\documentclass[sigconf]{acmart}

\usepackage{adjustbox}
\usepackage{balance}       
\usepackage{booktabs, multirow, makecell}
\usepackage[multiple,para]{footmisc} 
\usepackage{graphics}     
\usepackage{microtype}        
\usepackage{subfig}
\usepackage[flushleft]{threeparttable}

\renewcommand{\textrightarrow}{$\rightarrow$}

\copyrightyear{2018}
\acmYear{2018}
\setcopyright{acmcopyright}
\acmConference[DH'18]{2018 International Digital Health Conference}{April 23--26, 2018}{Lyon, France}
\acmBooktitle{DH'18: 2018 International Digital Health Conference, April 23--26, 2018, Lyon, France}
\acmPrice{15.00}
\acmDOI{10.1145/3194658.3194663}
\acmISBN{978-1-4503-6493-5/18/04}

\begin{document}
\title{Does Journaling Encourage Healthier Choices? \protect\\ Analyzing Healthy Eating Behaviors of Food Journalers}

\author{Palakorn Achananuparp}
\affiliation{%
  \institution{Singapore Management University}
  \city{Singapore} 
  \country{Singapore} 
}
\email{palakorna@smu.edu.sg}

\author{Ee-Peng Lim}
\affiliation{%
  \institution{Singapore Management University}
  \city{Singapore} 
  \country{Singapore} 
}
\email{eplim@smu.edu.sg}

\author{Vibhanshu Abhishek}
\affiliation{%
  \institution{Carnegie Mellon University}
  \city{Pittsburgh} 
  \state{PA} 
}
\email{vibs@andrew.cmu.edu}

\renewcommand{\shortauthors}{Achananuparp et al.}

\begin{abstract}

Past research has shown the benefits of food journaling in promoting mindful eating and healthier food choices. However, the links between journaling and healthy eating have not been thoroughly examined. Beyond caloric restriction, do journalers consistently and sufficiently consume healthful diets? How different are their eating habits compared to those of average consumers who tend to be less conscious about health? In this study, we analyze the healthy eating behaviors of active food journalers using data from MyFitnessPal. Surprisingly, our findings show that food journalers do not eat as healthily as they should despite their proclivity to health eating and their food choices resemble those of the general populace. Furthermore, we find that the journaling duration is only a marginal determinant of healthy eating outcomes and sociodemographic factors, such as gender and regions of residence, are much more predictive of healthy food choices.
\end{abstract}

%
%



\maketitle

\section{Introduction}
Recent progress in mobile and wearable technologies has provided individuals the means to routinely track data about themselves for self-knowledge and improvement. This self-tracking practice is also known as \textit{quantified self} or \textit{personal informatics}. In the domain of dietary self-monitoring, mobile food journal apps, such as MyFitnessPal (MFP hereafter), are one of the most popular tracking methods widely used by millions of people. Past research has suggested that food journaling is an effective intervention in weight loss programs \cite{Burke2012}. The act of journaling helps create increased in-the-moment awareness (mindfulness) and can encourage healthier choices \cite{Cordeiro2015a, Cordeiro2015}. Understanding how the journaling practice affects eating behaviors may provide useful insights for the designs of an effective population-wide health intervention. While there is growing evidence supporting the critical role of food journals in improving weight loss outcomes, little empirical work has been done to investigate the broader impacts of the journaling practice on the individuals' healthy eating behaviors. Especially, more evidence is needed to: (1) quantitatively compare long-term healthy eating habits of food journalers with other behavioral baselines, such as the general public and the dietary recommendations; and (2) measure the influences of the journaling practice on the healthy eating outcomes with respect to other factors. From a methodological perspective, it is also an opportunity to further explore the use of a large-scale self-tracking data in conjunction with offline data sources to answer these questions.


To address the research gaps, our study aims to assess the healthy eating behaviors of food journalers by analyzing public food diary entries of MFP users and comparing their eating behaviors to those of the general populace reported in other studies. Although people tend \textit{not} to perceive healthy eating the same ways as public-health experts \cite{Bisogni2009, rooksbyetal14chi}, a recent survey \cite{Cordeiro2015a} suggests that a vast number of food journalers (past and present) generally agree with experts about the notion of healthy eating, e.g., most believe that they should eat more fruits and vegetables, lean meat, and balanced diets. Thus, it is reasonable to assume that food journalers are more likely to achieve evidence-based healthy eating outcomes, as defined by public-health experts, than the average consumers who may be less informed about healthy eating. Next, we expect active food journalers, who are likely to develop a mindful eating habit, to consciously make healthy food choices and be less influenced by the sociodemographic biases. Specifically, we formulate the following research questions:

\textbf{RQ1:} \textit{Do active food journalers have healthier eating behaviors than the general populace?}
    
To investigate the effectiveness of journaling in encouraging healthier food choices, we aim to characterize the healthy eating behaviors of food journalers using the corresponding intakes from the dietary guidelines and the general populace as comparison data. If food journalers tend to (1) have higher intakes of healthy diets and lower intakes of unhealthy diets than the general populace and (2) consistently meet the recommended intakes per the dietary guidelines, then such findings may provide evidence supporting the notion that not only does journaling is linked to significant weight loss, but it also plays a significant role in individuals' healthier food choices.
    
\textbf{RQ2:} \textit{How do the eating behaviors of food journalers significantly differ across sociodemographic groups?}
    
\textit{Mindless eating} describes a situation where individuals are unaware of the influences exerted on their food choices by external factors, such as the environment and gender roles. It is generally associated with unhealthy eating habits and weight gains \cite{Wansink2006}. Here, we seek to further examine the relationship between the healthier food choices effect of journaling and the healthy eating behaviors of food journalers. Specifically, we expect food journalers of different sociodemographic backgrounds (e.g., gender, age, etc.) to be equally conscious of their food choices such that their healthy eating behavior is more homogeneous than that of the general populace. For example, male and female journalers should consume a comparable amount of fruits and vegetables. In other words, the difference in fruit and vegetable intakes between male and female journalers should not be statistically significant.
    
\textbf{RQ3:} \textit{To what extent does the journaling practice influence the eating behaviors of food journalers?}
    
Past research has shown that weight loss outcomes are proportional to the journaling practice. That is, individuals who are more active in recording their food journals tend to lose more weights than the less active journalers \cite{Burke2012}. As such, we aim to quantify the impacts of \textit{journaling duration} and \textit{persistence} on the healthy eating behaviors using regression analysis. If food journaling in fact encourages healthier food choices and mindful eating, we expect the journaling factors to have a higher influence on the eating behaviors than the sociodemographic factors. Such findings, together with those from the other research questions, will help demonstrate that food journaling is an effective healthy lifestyle intervention. On the other hand, unexpected results may provide us insights into the flaws of mobile food journals in affecting health behavior changes.
    


The main contributions of our work are as follows. \textit{Firstly}, we thoroughly examine the relationship between food journaling, sociodemographic factors, and a variety of healthy eating behaviors in a large population of nearly 10,000 active food journalers over a six-month period. \textit{Secondly}, we present our data preprocessing steps to automatically generate an analysis-ready dataset including: (1) identifying relevant foods and beverages from the annotated food diary entries; (2) extracting portion sizes of foods and beverages from the food diary entry text and the associated caloric value; and (3) normalizing the portion sizes of varying measurement units into standard nutritional units. \textit{Lastly}, our findings suggest, in contrast to past studies, that the journaling duration only plays a minor role in determining healthy eating behaviors, whereas several other factors, such as gender, journaling persistence, and regions of residence, are much more influential in determining the healthy eating behaviors than the journaling duration.

In what follows, we begin by reviewing related work on food journaling and data-driven approaches to health behavior assessment. Then, we introduce the healthy eating behaviors considered in the study. Next, we describe the dataset and data preprocessing steps. In the subsequent section, we define (1) the quantitative measures of the eating behaviors; and (2) the sociodemographic and journaling factors being studied. Finally, we present the findings, discuss their significance on the food journaling practice and future design implications, and conclude the paper.

\section{Related work}
\subsection{Personal informatics and food journaling }


Much research in personal informatics has focused on characterizing the use of tools and technologies to track one's own personal data for self-discovery and behavior change in a variety of domains \cite{rooksbyetal14chi, Epstein2015}. In particular, a few researchers have explored the use of mobile food journals and other online tools for dietary self-tracking in recent years \cite{Cordeiro2015, Cordeiro2015a, Epstein2016}. Cordeiro et al. identified several key challenges related to the journaling tools and practices, such as unreliable data and negative nudges \cite{Cordeiro2015}. To overcome tracking burden and promote mindful eating, Epstein et al. proposed a lightweight food journal \cite{Epstein2016}. Recently, Chung et al. \cite{Chung2017} studied the practice of food tracking amongst Instagram users and the role of social support on their healthy eating pursuit.

Our work is complementary to previous food journaling research \cite{Cordeiro2015a, Cordeiro2015, Epstein2015, Epstein2016}. While many studies aimed to qualitatively characterize various aspects of the self-tracking practices, few studies have taken a computational approach to examine the broader impacts of journaling on the healthy eating behaviors of food journalers. In this study, we analyze more than 1 million food diary entries to quantitatively assess the behavioral impacts of journaling.

\subsection{Using online data to assess health behaviors}


Data from online social media, quantified-self, and others have been used to study various aspects of health behaviors. First, a few studies \cite{culotta14chi, abbaretal15chi} analyzed mentions of foods in the Twitter network to track major public health issues. Next, Park et al. \cite{Park2016} investigated the impacts of user profile, fitness activity, and fitness network of Twitter users on the long-term engagement of fitness app users. Mejova et al. \cite{Mejova2015} analyzed food pictures shared by Instagram users to study the prevalence of obesity. Recently, a few studies have investigated the tasks of predicting diet compliance outcomes using MFP food diary data \cite{Weber2016} together with Twitter data \cite{DeChoudhury2017}.

Our work is highly relevant to \cite{DeChoudhury2017, Weber2016} in which the researchers constructed computational models to predict diet compliance success using different types of features, such as words \& food types identified from MFP diary entries and social and linguistic attributes extracted from the users' social media messages. While their studies particularly focused on caloric balance as the primary outcome, we examine a more comprehensive set of eating behaviors by using evidence-based healthy eating outcomes as the primary measures. Additionally, we investigate the role of sociodemographic and journaling factors, derived from the user profile and food diary data, in determining the healthy eating behaviors.

\section{Healthy eating behaviors}
We begin by introducing the evidence-based healthy eating outcomes categorized by the consumption of: (1) fruits and vegetables; (2) animal-based protein sources, such as red and processed meat, poultry, and fish; and (3) added sugars and sugary drinks. They are identified based on growing scientific evidence from several randomized controlled trials and meta-analysis about their associations with health benefits and risks. Together, they constitute dietary intakes commonly recommended by most dietary guidelines \cite{web:aha2016, who2003, web:cdc2015, web:who2015, web:acir, web:hsph2015}.


\subsection{Fruits and vegetables}\label{sec:methods:behaviors:fv}
High intake of fruits and vegetables (abbreviated as FV) provides a variety of long-term health benefits, such as lowering the risk of cardiovascular disease \cite{Wang2014} and cancers \cite{Latino-Martel2016}. On the other hand, low intakes of FV are associated with the increased prevalence of obesity and diabetes \cite{Mokdad2001}. Generally, a recommended daily FV intake for healthy adults is \textbf{at least 5 servings} \cite{Mokdad2001} or approximately 400 - 500 grams \cite{who2003, web:cdc2015}. Despite numerous health benefits, the consumption of fruits and vegetables has been persistently low in the US \cite{Blanck2008} and worldwide \cite{who2003} for decades. From 1994 - 2005, the average daily FV intake amongst Americans has decreased slightly from \textbf{3.43 servings} to \textbf{3.24 servings} while the percentage of people who met the recommended daily intake has remained unchanged at about \textbf{25\%} \cite{Blanck2008}. In 2015, the Centers for Disease Control and Prevention (CDC) reported that \textbf{less than 15\%} of Americans sufficiently met the recommendations \cite{web:cdc2015}.


\subsection{Red and processed meat, poultry, and fish}\label{sec:methods:behaviors:redmeat}
Growing evidence suggests that processed meat is carcinogenic while red meat is probably carcinogenic \cite{Latino-Martel2016}. High consumption of red and processed meat may increase mortality rates of type-2 diabetes \cite{Aune2009}, cardiovascular disease \cite{Song2016} and colorectal cancer \cite{Latino-Martel2016}. Replacing red and processed meat with healthier protein sources such as white meat (e.g., poultry and fish) \cite{Abete2014} may lower the risk of all-cause mortality. Most dietary guidelines recommend to limit a daily intake of red and processed meat to \textbf{1 serving} or approximately 65 - 75 grams \cite{web:acir}. Despite high associations with various health risks, red and processed meat still accounts for more than 50\% of meat consumed in the United States with almost \textbf{2.31 servings} per day \cite{Daniel2011}. Moreover, the consumption of healthier white meat, such as fish, amongst Americans is low. On average, most adults consumed \textbf{0.17 servings} (17.28 grams) of fish a day \cite{Papanikolaou2014}, 70.59\% lower than the recommended daily intake of \textbf{0.29 servings} (28.57 grams) \cite{web:aha2016}. 

\subsection{Added sugars and sugary drinks}\label{sec:methods:behaviors:sugar}
Added sugars are sugars not naturally occurring in foods and beverages. Asides from sweetening and adding extra calories, they provide no nutritional benefits. Sugary drinks (e.g., sodas/soft drinks, energy drinks, coffee drinks, and fruit juices) are the largest source (36\%) of added sugar intake in American diets \cite{web:health.gov}. According to the recent survey by Gallup \cite{web:gallup}, \textbf{48\%} of Americans drink at least 1 glass of soft drink on an average day. Ideally, sugary drink consumption should be avoided at all cost; or else an intake should be limited to \textbf{237 milliliters} a day \cite{web:hsph2015}. There is strong evidence linking high consumption of added sugars and sugary drinks to increased risks for obesity and type-2 diabetes, whereas simply lowering the intake of sugary drinks can reduce weight gain and decrease prevalence of obesity \cite{Hu2013}. Recent dietary guidelines suggest a maximum daily intake of added sugars to \textbf{25 grams} \cite{web:who2015}. However, an American adult consumes on average \textbf{70 grams} \cite{web:cdc2013} of added sugar per day, 180\% higher than the recommended amount.

\section{Data}
\subsection{Collecting and processing MFP data}\label{sec:methods:dataset}
We used a public food diary dataset\footnote{\url{http://bit.ly/2hNzRHT}} collected from 9,896 MFP users in March 2015 by Weber and Achananuparp \cite{Weber2016}. The dataset includes 71,715 unique food entries recorded over 1,919,024 meals from a 6-month recording period between October 2014 to March 2015. Each user recorded 59.3 days of diaries on average (S.D. = 54.6, median = 42). The majority of users were able to achieve their daily caloric goals \cite{Weber2016}. Users who recorded at least 7 days of diary entries were treated as active users (N=8,381; 85.69\% of all users in the dataset). Each food diary text was automatically annotated with categorical information describing its composition (e.g., food groups) and cooking method in the previous work \cite{Weber2016}. In addition, basic nutritional facts (e.g., calories, protein, sugar) for each food entry were available. Next, we performed data cleaning by removing 207 outlying diary entries (0.04\% of total data) whose total daily calories are: (1) greater than 6,000 kcal or 2 standard deviations away from the mean daily calories; or (2) lower than or equal to zero kcal.




For each user in the dataset, we further collected personal information from their user profile page. In total, the profile pages of 8,794 users are publicly accessible. Next, we categorized the profile attributes such as age, geographical location, and friend list into the following groups: Age group (young adults age 18-44 years old and old adults age 45 and above), social connection quartiles (Q1: 0-6 friends, Q2: 7-18 friends, Q3: 19-41 friends, and Q4: 42 or more friends), regions of residence at a global level (US and Non-US), and regions of residence within the United States as per the US census classification (Northeast, South, Midwest, and West). As we can see in Table \ref{tbl:demo}, the majority of users are female (82.01\%) and young adults 18 - 44 of age (79.80\%). Next, more than 50\% of users have less than 19 users in their friend list. Most users reside in the United States (70.88\%) while the rest of the users live outside the US. Amongst US users, 33.76\% live in the Southern states, whereas 18.37\% live in the Northeastern states. The geographical distribution of MFP users is representative of the population distribution across US regions \cite{web:census}.


\begin{table}[htb]
\centering
\caption{Sociodemographic distributions}
\label{tbl:demo}
\scalebox{0.8}{
\begin{tabular}{@{}llcc@{}}
\toprule
 &  & Count & \% \\ \midrule
Gender & Female & 7,212 & 82.01\% \\
 & Male & 1,582 & 17.99\% \\ \midrule
Age group & 18-44 & 7,015 & 79.80\% \\
 & 45+ & 1,776 & 20.20\% \\ \midrule
Social connection & Q1 (0-6) & 2,681 & 30.49\% \\
 & Q2 (7-18) & 2,170 & 24.68\% \\
 & Q3 (19-41) & 2,040 & 23.20\% \\
 & Q4 (42+) & 1,903 & 21.64\% \\ \midrule
Global region & US & 6,233 & 70.88\% \\
 & Non-US & 2561 & 29.12\% \\ \midrule
US region & South & 2,104 & 33.76\% \\
 & Midwest & 1,593 & 25.56\% \\
 & West & 1,391 & 22.32\% \\
 & Northeast & 1,145 & 18.37\% \\ \bottomrule
\end{tabular}}
\end{table}

\begin{table}[thpb]
\centering
\caption{Amounts of food equivalent to 1 serving size}
\label{tbl:conversion}
\scalebox{0.8}{
\begin{tabular}{@{}lll@{}}
\toprule
 & Grams & kcal \\ \midrule
Fruit & 150 & 75 \\
Vegetable & 75 & 40 \\
Red meat & 65 & 160 \\
Poultry & 80 & 160 \\
Fish & 100 & 160 \\ \bottomrule
\end{tabular}}
\end{table}

\subsection{Normalizing and extracting portion sizes}\label{sec:methods:normalizing}

Each diary entry contains a free-text description of the food item (i.e., name and portion size) and nutritional facts, such as calories, protein, sugar. When recording a diary entry, MFP users can choose to enter the amount of food and beverage in various measurement units. For example, the measurement units for \textit{apple} may include weights (grams, ounces, pounds, and kilograms), volumes (cup), physical sizes (small, medium, and large), and nutritional units (servings). To make it possible to directly compare the dietary intakes, we apply the following steps to normalize the portion sizes. First, we extracted quantity and unit from the diary entry text using regular expressions and convert weights and volumes from other measurement systems to grams and milliliters, respectively. Then, we used the annotated categories to identify specific types of diets. First, for solid foods containing only a category of fruit, vegetable, red meat, poultry, or fish, we determined standard serving sizes using the conversions \cite{web:eatforhealth} in Table \ref{tbl:conversion}. For example, 100 grams of \textit{apple} is equal to 0.67 serving of fruit. If a gram-equivalent weight cannot be found in the diary entry text, the amount of corresponding calories in the diary entry was used for the conversion instead. E.g., a \textit{320-kcal tuna} contains 2 servings of fish. Next, for composite foods typically served with grains and other types of ingredients, we subtract 240 (an average kcal for 1 regular serving of grain-based foods) from the total calories and calculate the serving sizes using the corresponding caloric values in Table \ref{tbl:conversion}. With these steps, we can identify 1 serving of fish from \textit{a 400-kcal tuna sandwich}.

\subsection{Caloric intake patterns}\label{sec:methods:data_char}

\begin{figure}[thpb]
\begin{tabular}{cccc}
\subfloat[Caloric intake by month]{
\includegraphics[width = 1.6in]{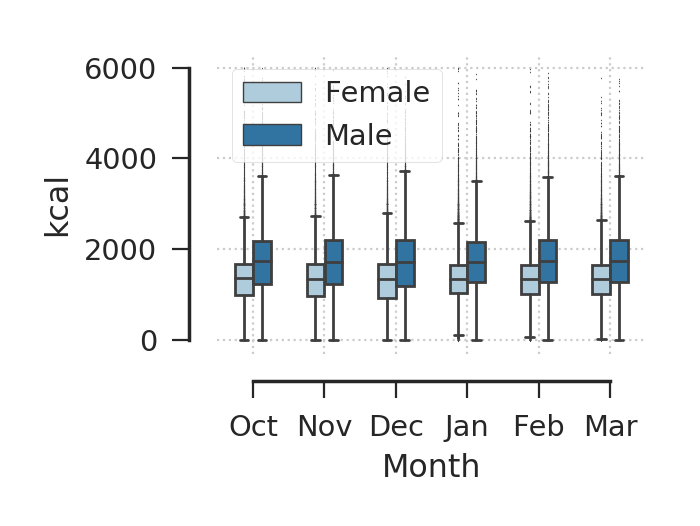}
\label{fig:box_cal_monthly}} &
\subfloat[Caloric intake by weekday]{
\includegraphics[width = 1.6in]{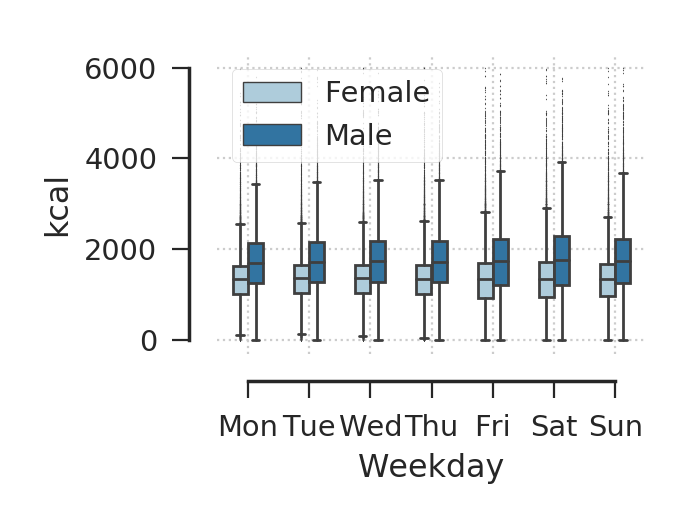}
\label{fig:box_cal_weekday}}
\end{tabular}
\caption{Average daily caloric intakes over time (kcal)}
\label{fig:data_char}
\end{figure}

Figure \ref{fig:data_char} displays the average daily caloric intakes over time. Overall, the average daily caloric intakes, as seen in Figure \ref{fig:data_char}\subref{fig:box_cal_monthly} are around 1,700 kcal and 1,300 kcal for male and female users, respectively. As expected, these are much lower than the estimated calorie requirements of an average adult (2,000 - 2,200 kcal), suggesting that the users were likely dieting. Furthermore, we observe the largest interquartile range of the daily caloric intakes in December, followed by the smallest in January, a possible effect of a new year's weight-loss resolution. Lastly, we compare the daily calorie intakes between weekdays in Figure \ref{fig:data_char}\subref{fig:box_cal_weekday}. As can be seen, the daily caloric intakes follow the weekday-weekend lifestyle pattern, trending slightly upward from Monday before reaching the largest interquartile range on Saturday.

\section{Methods}

\subsection{Behavioral measures}
Based on the eating behaviors introduced previously, we define the following measures to quantify the behavioral outcomes for each MFP user:

\textbf{Journaling behaviors:} Two journaling behaviors are defined to represent the journaling duration and persistence. First, \textit{recording days} is defined as a number of days the user records food diary entries. Second, \textit{normalized lapsing frequency} is defined as a fraction of days the user temporarily stops recording any diary entries with respect to her journaling lifetime (recording days + lapsing days).


\textbf{Eating behaviors:} To measure the energy intake from foods and beverages, we compute \textit{median daily caloric intake} (in kcal). Next, for each diet type, we compute \textit{median daily intake} (in servings for foods, grams for added sugars, and milliliters for drinks) and \textit{normalized intake frequency} (\% of days in which the diets were consumed). We used the following criteria to identify specific diet types from the annotated food diary entries. First, fruits and vegetables are selected from entries tagged with \textit{fruit} and \textit{vegetable} categories. Next, red and processed meats are chosen from entries tagged with \textit{beef}, \textit{pork}, \textit{lamb}, \textit{game} (meat from wild animals), \textit{sausage}, and \textit{meatball} sub-categories. Furthermore, poultry and fish are identified from entries tagged with \textit{poultry} and \textit{fish} sub-categories, respectively. Next, added sugars are non-zero sugar content entries tagged with the following categories and sub-categories: \textit{beverage}, \textit{dessert}, \textit{snack}, \textit{condiment}, and \textit{dairy product}. Lastly, sugary drinks include any non-zero sugar content entries tagged with \textit{beverage} category, whereas soft drinks included entries tagged with \textit{soft drink} subcategory. This results in 14 diet-specific behavioral measures.


\begin{figure*}[ht]
\centering
\scalebox{0.8}{
\begin{tabular}{cccc}
\subfloat[FV intake]{
\includegraphics[width = 1.6in]{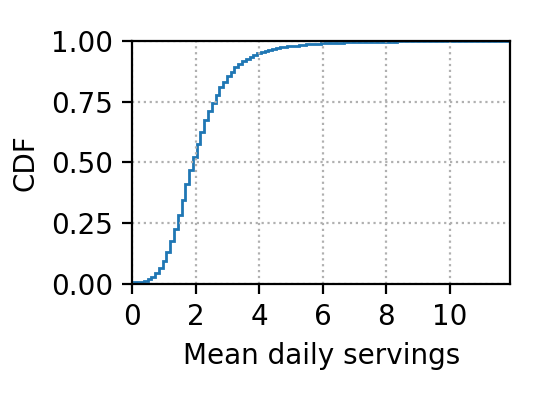}
\label{fig:cdf_fv_servs}} &
\subfloat[Meat intake]{
\includegraphics[width = 1.6in]{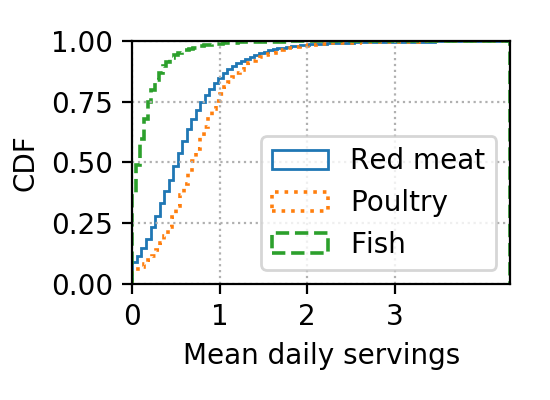}
\label{fig:cdf_pro_servs}} &
\subfloat[Added sugar intake]{
\includegraphics[width = 1.6in]{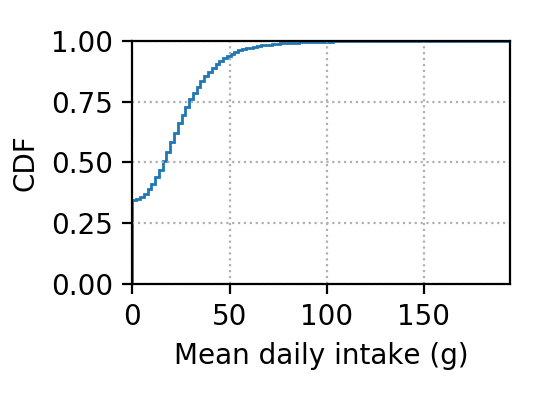}
\label{fig:cdf_sugar_servs}} &
\subfloat[Sugary drink intake]{
\includegraphics[width = 1.6in]{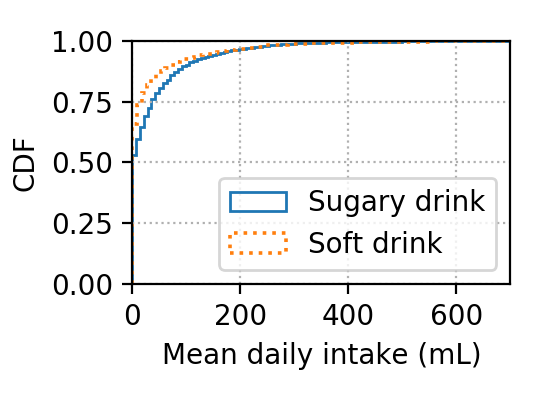}
\label{fig:cdf_drink_servs}} \\
\subfloat[FV frequency]{
\includegraphics[width = 1.6in]{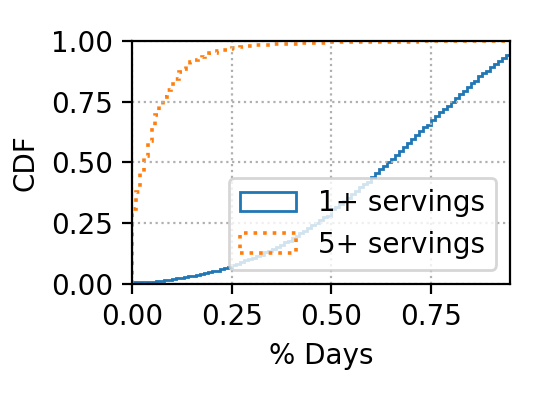}
\label{fig:cdf_fv_freq}} &
\subfloat[Meat frequency]{
\includegraphics[width =1.6in]{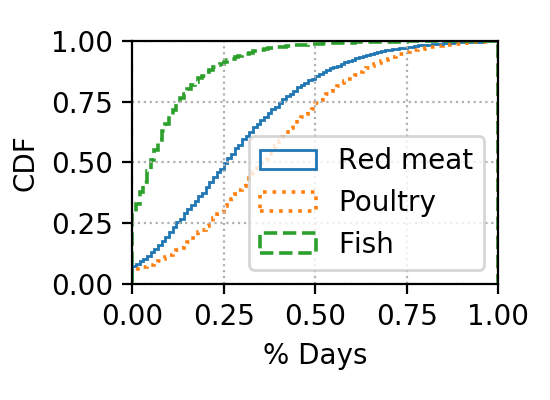}
\label{fig:cdf_pro_freq}} &
\subfloat[Added sugar frequency]{
\includegraphics[width = 1.6in]{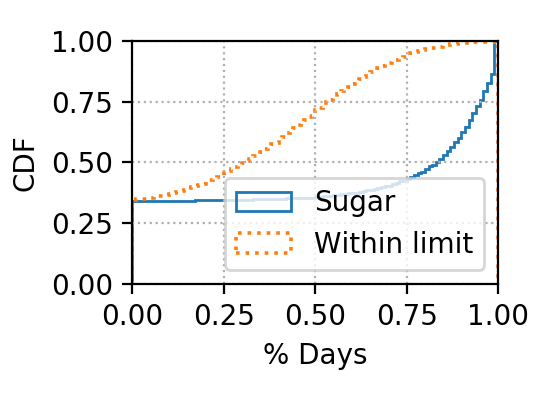}
\label{fig:cdf_sugar_freq}} &
\subfloat[Sugary drink frequency]{
\includegraphics[width = 1.6in]{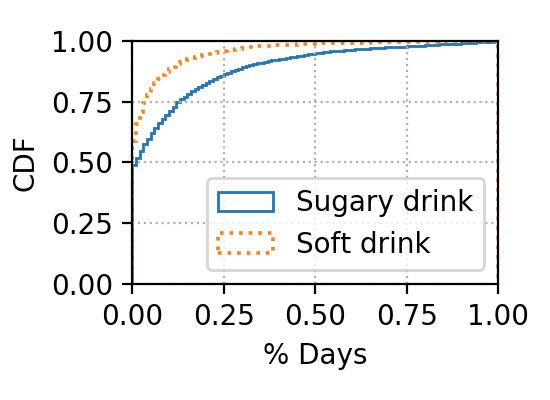}
\label{fig:cdf_drink_freq}}
\end{tabular}}
\caption{Average daily servings and frequencies}
\label{fig:cdf}
\end{figure*}

\subsection{Sociodemographic and journaling factors}
Now, we introduce the following factors known to be associated with specific healthy eating behaviors below.


\textbf{Gender and age:} Gender and age have been found as contributors of behavioral differences in many dietary behavior studies. For example, women tended to to eat more healthy diets than men, e.g., consuming more fruits and fiber \cite{Blanck2008}. High consumption of meat and red meat was closely associated with being male \cite{Daniel2011, Gossard2003}. Men also consumed more sugary drinks than women \cite{Bleich2008}. Next, as people get older, they tended to eat less and changed their eating behaviors. For instance, compared to younger adults, older adults were more likely to consume more fruits and vegetables \cite{Blanck2008}, less red meat as well as all meat \cite{Daniel2011, Gossard2003} but more fish \cite{Daniel2011}, and less sugary drink \cite{Bleich2008}. In the mean time, younger adults tended to consume more poultry \cite{Daniel2011}. In this work, we investigate (1) the behavioral differences between genders (male and female), age groups (young and old adults) in RQ2 and (2) the influences of the gender and age factors on all behavioral measures in RQ3.
    
\textbf{Social connections:} Social ties have shown to have both the positive and negative influences on health behaviors \cite{Umberson2010}. For instance, daily consumption of fruits and vegetables \cite{Nieminen2013} and overall weight loss \cite{Hwang2010} were associated with high levels of social support. However, having obese friends substantially increased one's own risk of obesity \cite{Christakis2007}. In this study, we explore (1) the behavioral differences between social connection quartiles in RQ2 and (3) the influence of social connections on all behavioral measures in RQ3.
    
\textbf{Regions of residence:} Differences in eating behaviors were observed amongst Americans in different regions, which could be attributed to \textit{social norms} and \textit{environmental contexts}. People in the Northeast and the West were more likely to consume more fruits and vegetables than those in the Midwest and the South \cite{Blanck2008}. Next, beef consumption was highest in Midwest and lowest in the South \cite{Gossard2003}. In addition, people in the Midwest and the Northeast consumed more sugary drinks than those in the South and the West \cite{Park2015}. Particularly, the consumption of soft drinks was the highest in the Northeast, compared to other regions \cite{Park2015}. In this work, we examine (1) the behavioral differences between regions of residence within the United States (Northeast, South, Midwest, and West) as well as global regions of residence (US vs. non-US) in RQ2 and (2) their influences on all behavioral measures in RQ3.
    
\textbf{Journaling:} Food journaling can promote mindful eating \cite{Cordeiro2015a}, leading to weight loss \cite{Hollis2008}, and healthier behaviors \cite{Epstein2016}. Furthermore, the health benefits were greater amongst highly active journalers \cite{Burke2012}. In this work, we explore the influence of the journaling behaviors, i.e., recording days and lapsing frequency, on all caloric and diet-specific behavioral measures in RQ3.

\section{Results}
\subsection{RQ1: Distributions of the healthy diet intakes}\label{sec:results:cdf} 
We first examine the average daily intake and the intake frequency of each diet type over the 6-month period. For each food type, we calculated the cumulative distribution function (CDF) of the mean daily servings and the percentage of days consumed for each user. All CDF plots are shown in Figure \ref{fig:cdf}.

\textbf{Fruits and vegetables:} As shown in Figure \ref{fig:cdf}\subref{fig:cdf_fv_servs}, most food journalers do not consume a sufficient amount of fruits and vegetables on a daily basis. Especially, the average daily FV intake of food journalers is much lower than that of the general populace. In particular, 50\% of users consume up to \textbf{1.97 servings} of FV per day, whereas only \textbf{2.18\%} of users (N=183) manage to meet the recommended daily intake of 5 servings. In terms of the frequency of intake, shown in Figure \ref{fig:cdf}\subref{fig:cdf_fv_freq}, 50\% of users consume fruits and vegetables up to 64.96\% of the time. Furthermore, \textbf{less than 1\%} of users (N=6) are able to meet the recommended intake at least 80\% of the time. This is very surprising since fruits and vegetables should be an essential part of any healthy diets, particularly amongst people trying to lose weight and live a healthier lifestyle.


\textbf{Red and processed meat, poultry, fish:} Unlike typical American consumers, food journalers consume higher portions of healthy protein sources, especially poultry, than unhealthy red and processed meat. As shown in Figure \ref{fig:cdf}\subref{fig:cdf_pro_servs}, 50\% of users consume up to \textbf{0.53 servings} of red and processed meat and \textbf{0.72 servings} of poultry per day. Moreover, the average intakes are within the recommended daily limit of 2 servings. Next, poultry is also consumed more frequently than red and processed meat. According to Figure \ref{fig:cdf}\subref{fig:cdf_pro_freq}, 50\% of users consume red meat and poultry up to \textbf{26\%} and \textbf{36\%} of the time, respectively. However, the average daily fish intake is much lower than that of other protein sources. Similar to average consumers, food journalers consume much fewer servings of fish per day than the recommended daily intake. Particularly, 50\% of users consume \textbf{0.09 servings} of fish a day, whereas only \textbf{15\%} of users (N=1,444) meet the recommendation as shown in Figure \ref{fig:cdf}\subref{fig:cdf_pro_servs}.

\textbf{Added sugar and sugary drinks:} The average daily added sugar intake amongst food journalers is much lower than that of the general populace. According to Figure \ref{fig:cdf}\subref{fig:cdf_sugar_servs}, 50\% of users consume up to \textbf{16 grams} of added sugar per day. However, the top 25\% of users consume higher amount of daily added sugar intake (\textbf{29.83 grams}) than the recommended limits. Furthermore, the consumption occurs very frequently. As shown in Figure \ref{fig:cdf}\subref{fig:cdf_sugar_freq}, 50\% of users consume added sugars up to \textbf{83\%} of the time, whereas the consumption within the recommended limit only accounts for up to \textbf{30\%} of the time. Next, the overall consumption of sugary drinks and soft drinks amongst food journalers is comparable to that of the general populace. As shown in Figure \ref{fig:cdf}\subref{fig:cdf_drink_servs}, \textbf{53.25\%} and \textbf{60.66\%} of users do not consume any sugary/soft drinks, respectively. Amongst drinkers, the average daily intakes for both sugary and soft drinks are less than 1 glass. However, on any drinking days, the daily intake exceeds the limit most of the time.

\textbf{Key insights:} Unexpectedly, we find that the healthy eating behaviors of active food journalers do not differ much from those of the general populace in several areas. Given that our population is highly skewed toward females, who tend to be health-conscious, the numbers of healthy eating lapses observed are even more surprising. For example, a vast majority of food journalers did not eat enough fruits and vegetables and fish as per the dietary guidelines. Next, their sugary and soft drink consumption was about the same level as the general populace, which is a worrying trend. On a positive note, the consumption of red and processed meat and added sugars is lower in food journalers than the general populace.


\begin{table}[tpb]
\centering
\caption{Behavioral differences between groups.}
\label{tbl:stat-tests}
\begin{adjustbox}{max width=\columnwidth}
\begin{threeparttable}
\begin{tabular}{@{}lcccccc@{}}
\toprule
\multirow{2}{*}{Behavioral measure} & \multirow{2}{*}{Gender} & \multirow{2}{*}{Age} & \multirow{2}{*}{Friends} & \multicolumn{2}{c}{Region} \\ \cmidrule(l){5-6} 
 & & & \multicolumn{1}{c}{} & US & Global \\ \midrule
Recording days & Male\textsuperscript{**} & 45+\textsuperscript{**} & Q4\textsuperscript{**} & Northeast\textsuperscript{**} & Non-US\textsuperscript{*} \\
Lapsing frequency & Female\textsuperscript{**} & 18-44\textsuperscript{**} & Q1\textsuperscript{**} & & \\ \midrule
Caloric intake & Male\textsuperscript{**} & 18-44\textsuperscript{*} & Q4\textsuperscript{**} & & Non-US\textsuperscript{**} \\ \midrule
FV intake\textsuperscript{**} & Male\textsuperscript{**} & & Q4\textsuperscript{*} & West\textsuperscript{**} & Non-US\textsuperscript{**} \\
FV frequency & Female\textsuperscript{**} & & Q4\textsuperscript{**} & West\textsuperscript{**} & Non-US\textsuperscript{**} \\ \midrule
Red meat intake & Male\textsuperscript{**} & 45+\textsuperscript{**} & & Midwest\textsuperscript{**} & Non-US\textsuperscript{**} \\
Poultry intake & Male\textsuperscript{**} & 18-44\textsuperscript{**} & Q4\textsuperscript{*} & South\textsuperscript{**} & US\textsuperscript{**} \\
Fish intake & & & & & US\textsuperscript{**} \\
Red meat frequency & Male\textsuperscript{**} & 45+\textsuperscript{**} & Q4\textsuperscript{*} & Midwest\textsuperscript{**} & Non-US\textsuperscript{**} \\
Poultry frequency & Male\textsuperscript{**} & 18-44\textsuperscript{**} & Q4\textsuperscript{**} & South\textsuperscript{*} & US\textsuperscript{**} \\
Fish frequency & Male\textsuperscript{**} & 45+\textsuperscript{**} & & & Non-US\textsuperscript{**} \\ \midrule
Added sugar intake & & 18-44\textsuperscript{*} & Q1\textsuperscript{**} & & US\textsuperscript{**} \\
Sugary drink intake & & & & & & \\
Soft drink intake & Male\textsuperscript{**} & & & Midwest\textsuperscript{*} & US\textsuperscript{*} \\
Added sugar frequency & Female\textsuperscript{**} & & Q1\textsuperscript{**} &  & US\textsuperscript{**} \\
Sugary drink frequency & Female\textsuperscript{*} & 18-44\textsuperscript{**} & Q1\textsuperscript{**} & & \\
Soft drink frequency & & 18-44\textsuperscript{**} & Q1\textsuperscript{**} & South\textsuperscript{**} & US\textsuperscript{**} \\ \bottomrule
\end{tabular}
\begin{tablenotes}
\small
\item Significance thresholds are * p<0.05 and ** p<0.01. Caloric intake is in kcal. Food intakes are in numbers of servings. Sugar intake is in grams. Drink intakes are in mL. Intake frequencies are in percentages of days. Cell values represent groups with the highest median. Non-statistically significant results are omitted.
\end{tablenotes}
\end{threeparttable}
\end{adjustbox}
\end{table}

\begin{table}[thpb]
\centering
\caption{Dunn's multiple comparisons for different social connections and US regions.}
\label{tbl:posthoc}
\begin{adjustbox}{max width=\columnwidth}
\begin{threeparttable}
\begin{tabular}{@{}lll@{}}
\toprule
Behavioral measure & Friends & US regions \\ \midrule
Recording days & 1\textbf{2}\textsuperscript{*}, 1\textbf{3}\textsuperscript{**}, 1\textbf{4}\textsuperscript{**}, 2\textbf{3}\textsuperscript{**}, 2\textbf{4}\textsuperscript{**}, 3\textbf{4}\textsuperscript{**} & \textbf{N}S\textsuperscript{*}, S\textbf{M}\textsuperscript{*} \\
Lapsing frequency & \textbf{1}4\textsuperscript{**}, \textbf{2}4\textsuperscript{**}, \textbf{3}4\textsuperscript{**} 
&  \\ \midrule
Caloric intake & 1\textbf{4}\textsuperscript{*}, 2\textbf{4}\textsuperscript{**}, 3\textbf{4}\textsuperscript{**} 
&  \\ \midrule
FV intake & 2\textbf{4}\textsuperscript{*} 
& S\textbf{M}\textsuperscript{*}, M\textbf{W}\textsuperscript{**} \\
FV frequency & 1\textbf{4}\textsuperscript{**}, 2\textbf{3}\textsuperscript{*}, 2\textbf{4}\textsuperscript{**} 
& \textbf{N}S\textsuperscript{**}, \textbf{N}M\textsuperscript{**}, S\textbf{W}\textsuperscript{**}, M\textbf{W}\textsuperscript{**} \\
Red meat intake &  & N\textbf{S}\textsuperscript{*}, N\textbf{M}\textsuperscript{**} \\
Poultry intake & 3\textbf{4}\textsuperscript{**} 
& \textbf{N}M\textsuperscript{*}, \textbf{S}M\textsuperscript{*} \\
Fish intake &  &  \\
Red meat frequency & 1\textbf{4}\textsuperscript{*} 
& N\textbf{S}\textsuperscript{**}, N\textbf{M}\textsuperscript{**}, N\textbf{W}\textsuperscript{**} \\
Poultry frequency & 2\textbf{4}\textsuperscript{**}, 3\textbf{4}\textsuperscript{**} 
& \textbf{S}M\textsuperscript{*} \\
Fish frequency &  &  \\ \midrule
Added sugar intake & \textbf{1}2\textsuperscript{**}, \textbf{1}3\textsuperscript{**}, \textbf{1}4\textsuperscript{**}, \textbf{2}4\textsuperscript{**}, \textbf{3}4\textsuperscript{**} &  \\
Sugary drink intake &  &  \\
Soft drink intake &  &  \\
Added sugar frequency & \textbf{1}2\textsuperscript{**}, \textbf{1}3\textsuperscript{**}, \textbf{1}4\textsuperscript{**}, \textbf{2}4\textsuperscript{**}, \textbf{3}4\textsuperscript{**} &  \\
Sugary drink frequency & \textbf{1}3\textsuperscript{**}, \textbf{1}4\textsuperscript{**}, \textbf{2}4\textsuperscript{**} &  \\
Soft drink frequency & \textbf{1}4\textsuperscript{**} 
& N\textbf{S}\textsuperscript{**}, \textbf{S}M\textsuperscript{*}, \textbf{S}W\textsuperscript{**} \\ \bottomrule
\end{tabular}
\begin{tablenotes}
\small
\item Significant thresholds are * p<0.05 and ** p<0.01. Social connection groups are abbreviated as 1: Q1, 2: Q2, 3: Q3, and 4: Q4. Region groups are abbreviated as N: Northeast, S: South, M: Midwest, and W: West. Comparisons are denoted as a pair of letters, e.g., 12 represents a comparison of two social connection groups Q1 and Q2. Groups with a higher rank sum are in bold. Non-statistically significant results are omitted.
\end{tablenotes}
\end{threeparttable}
\end{adjustbox}
\end{table}

\subsection{RQ2: Behavioral differences}\label{sec:results:diff}
In this section, we present the differences in eating behaviors of food journalers across sociodemographic groups. Comparisons of behavioral measures, shown in Table \ref{tbl:stat-tests}, were performed by (1) Mann-Whitney U test for genders, age groups, and global regions; and (2) Kruskal-Wallis H test for social connections and US regions. Dunn's multiple comparisons, shown in Table \ref{tbl:posthoc}, were performed for social connections and US regions for post-hoc tests.

\textbf{Journaling and caloric intakes:} First, significantly higher median recording days are observed amongst the following groups than other groups: males (p<0.01), older adults (p<0.01), largest social connections (p<0.01), residing in the Northeast (p<0.01), and residing outside the US (p<0.05) as shown in Table \ref{tbl:stat-tests}; pairwise differences are significant for all social connection pairs and the median recording days of the Northeast and the Midwest are significantly higher than that of the South as displayed in Table \ref{tbl:posthoc}. Second, median lapsing frequency are significantly higher amongst users in the following groups: females (p<0.01), younger adults (0.01), and small connections (0.01); pairwise differences are significant between smaller social connections (Q1, Q2, and Q3) vs. the largest (Q4). Significantly higher median caloric intakes are observed amongst the following groups than other groups: males (p<0.01), younger adults (p<0.05), largest social connections (p<0.01), and residing outside the US (p<0.01); pairwise differences are significant between smaller social connections (Q1, Q2, and Q3) vs. the largest (Q4).

\textbf{Fruits and vegetables:} Compared to the general populace, the differences in FV consumption of food journalers are associated with similar sociodemographic groups (gender, social connections, and regions of residence). First, females consume fruits and vegetables more frequently than males (p<0.01). Though, unlike the general populace, males have a significantly higher median FV intake (p<0.01). Next, the differences in FV intake and intake frequency are significant between groups with different social connections (p<0.05 and p<0.01, respectively); pairwise differences are significant between (Q2 vs. Q4) for FV intake and (Q1 vs. Q4), (Q2 vs. Q3), and (Q2 vs. Q4) for intake frequency. Moreover, significantly higher median FV intake and frequency are observed amongst those in the West (p<0.01); pairwise intake differences are significant between (South vs. West) and (Midwest vs. West), whereas pairwise frequency differences are significant between all region pairs except (Northeast vs. West) and (South vs. Midwest). In contrast to the general populace, there are no behavioral differences between age groups for FV consumption.

\begin{table*}[ht]
\centering
\caption{Coefficients ($\beta$) of 17 OLS regression models.}
\label{tbl:regression}
\resizebox{\textwidth}{!}{%
\begin{threeparttable}
\begin{tabular}{lcccccccccc}
\toprule
Predicted variable & Male & Age & \makecell{Log\\(Friends)} & Northeast & South & West & Non-US & \makecell{Recording\\days} & \makecell{Lapsing\\frequency} & \makecell{Adjusted\\R\textsuperscript{2}} \\ \midrule
Recording days & $9.470^{**}$ & $0.653^{**}$ & $7.725^{**}$ & 2.837 & $-4.292^{*}$ & -2.939 & 2.291 & & & 0.055 \\
Lapsing frequency & $-0.047^{**}$ & $-0.004^{**}$ & $-0.019^{**}$ & -0.003 & 0.011 & 0.011 & -0.007 & & & 0.026 \\ \midrule
Caloric intake & $343.71^{**}$ & $-3.458^{**}$ & $7.498^{*}$ & -14.235 & -19.207 & 19.598 & $84.805^{**}$ & $1.163^{**}$ & $-312.951^{**}$ & 0.219 \\ \midrule
FV intake & ${0.203^{**}}$ & -0.002 & 0.011 & 0.071 & 0.04 & $0.107^{*}$ & $0.198^{**}$ & -0.001 & $-0.268^{**}$ & 0.012 \\
FV frequency & $-0.072^{**}$ & $0.001^{*}$ & 0.004 & $0.032^{**}$ & 0.002 & $0.036^{**}$ & $0.104^{**}$ & $0.0002^{*}$ & $-0.107^{**}$ & 0.069 \\ \midrule
Red meat intake & $0.228^{**}$ & $0.004^{**}$ & $0.012^{*}$ & $-0.083^{**}$ & $-0.041^{*}$ & -0.031 & 0.016 & $-0.0004^{*}$ & -0.014 & 0.040 \\
Poultry intake & $0.235^{**}$ & $-0.006^{**}$ & $0.010^{*}$ & $0.051^{*}$ & $0.055^{**}$ & $0.039^{*}$ & $-0.121^{**}$ & 0.0001 & 0.0056 & 0.051 \\
Fish intake & 0.006 & 0.0003 & -0.0003 & 0.012 & 0.005 & 0.005 & $0.055^{**}$ & $-0.0002^{**}$ & -0.0193 & 0.011 \\
Red meat frequency & $0.057^{**}$ & $0.002^{**}$ & $0.006^{**}$ & $-0.037^{**}$ & -0.003 & -0.008 & $0.018^{*}$ & $0.0002^{**}$ & $-0.0561^{**}$ & 0.062 \\
Poultry frequency & $0.055^{**}$ & $-0.001^{**}$ & $0.004^{*}$ & 0.007 & $0.024^{**}$ & 0.007 & $-0.043^{**}$ & $0.0003^{**}$ & $-0.0718^{**}$ & 0.052 \\
Fish frequency & $0.009^{*}$ & $0.0003^{*}$ & -0.001 & 0.006 & 0.007 & 0.004 & $0.047^{**}$ & -0.0004 & $-.0309^{**}$ & 0.029 \\ \midrule
Added sugar intake & 0.291 & $-0.133^{**}$ & $-1.454^{**}$ & -0.407 & -0.238 & -0.439 & $-3.701^{**}$ & $-0.015^{*}$ & $-7.142^{**}$ & 0.028 \\
Sugary drink intake & -3.066 & 0.117 & -0.821 & $5.548^{*}$ & $5.045^{*}$ & 1.151 & $7.229^{**}$ & -0.04 & -6.96 & 0.002 \\
Soft drink intake & $6.714^{**}$ & 0.143 & 0.595 & -3.613 & -1.326 & $-6.412^{*}$ & 0.23 & -0.042 & -1.209 & 0.002 \\
Added sugar frequency & $-0.056^{**}$ & $-0.001^{*}$ & $-0.044^{**}$ & 0.008 & 0.013 & 0.021 & $-0.049^{**}$ & $-0.0008^{**}$ & $-0.169^{**}$ & 0.032 \\
Sugary drink frequency & $-0.014^{*}$ & $-0.001^{*}$ & $-0.006^{**}$ & 0.012 & 0.012 & 0.007 & $0.022^{**}$ & -0.0001 & $-0.0293^{**}$ & 0.005 \\
Soft drink frequency & $0.009^{**}$ & $-0.0003^{*}$ & 0.0001 & -0.003 & $0.009^{*}$ & -0.006 & $0.017^{**}$ & -0.0003 & $-0.013^{*}$ & 0.008 \\ \bottomrule
\end{tabular}
\begin{tablenotes}
\small
\item Caloric intake is in kcal. Food intakes are in numbers of servings. Sugar intake is in grams. Drink intakes are in mL. Intake frequencies are in percentages of days. Intercept terms are omitted. Significance thresholds are * p<0.05 and ** p<0.01.
\end{tablenotes}
\end{threeparttable}
}
\end{table*}

\textbf{Red and processed meat, poultry, fish:} Similar to the general populace, the differences in meat consumption amongst the MFP populace are associated with gender, age, and regions of residence. First, males have a significantly higher median intake (p<0.01) and frequency (p<0.01) than females, for all protein sources except fish. Second, older adults have a significantly higher fish intake frequency (p<0.01). In addition, poultry consumption (median intake and frequency) in younger adults is significantly higher (p<0.01). Next, differences in red and processed meat intake frequency are significant amongst regions of residence (p<0.01). Users in the Midwest generally have the highest intake and intake frequency; pairwise intake differences are significant between (Northeast vs. South) and (Northeast vs. Midwest), whereas pairwise frequency differences are significant between the followings: (Northeast vs. South), (Northeast vs. Midwest), and (Northeast vs. West). As opposed to the general populace, older adults have a significantly higher median red and processed meat intake (p<0.01) and frequency (p<0.01) than younger adults.

\textbf{Added sugars and sugary drinks:} Lastly, the differences in added sugar and sugary drink consumption of food journalers are associated with gender, age, and regions of residence -- reflecting the overall differences in the general populace. First, males have a significantly higher median intake of soft drink (p<0.01) than females; however, in contrast to the general populace, females have significantly higher intake frequencies of added sugars (p<0.01) and sugary drink (p<0.05). Second, the median intake of added sugars is significantly higher in younger adults (p<0.05). Furthermore, younger adults have significantly higher intake frequencies of sugary drinks (p<0.01) and soft drinks (p<0.01), than older adults. In contrast to the general populace, we find that food journalers in the Midwest and the South tend to consume more soft drinks than those in the Northeast and the West. Specifically, users in the Midwest have a significantly higher median soft drink intake (p<0.05), whereas those in the South have a significantly higher soft drink intake frequency (p<0.01) than other regions; pairwise differences are significant for the South vs. other regions for the soft drink intake frequency, whereas there are no pairwise differences between regions for the median soft drink intake.

\textbf{Key insights:} Overall, food journalers who are male, 45 years or older, and have the largest social network tend to have significantly longer journaling duration and more persistent in recording food journals than others. Next, the healthy eating behaviors of food journalers within the sociodemographic groups are not as homogeneous as we initially expected. As can be seen in Table \ref{tbl:stat-tests}, 58 of 85 (68\%) behavioral differences are statistically significant. Furthermore, the differences in eating behaviors across sociodemographic groups are, in many cases, fairly similar to those naturally observable in the general populace. Specifically, both positive and negative eating behaviors occurred within the expected sociodemographic groups, e.g., high intake frequency of fruits and vegetables in females, high added sugar intake in younger adults, and more healthy eating behaviors in users with larger social connections, etc. The results are quite interesting as they may suggest that food journalers were not being as mindful of their healthier food choices as they should have been. We will further investigate the influences of sociodemographic and journaling factors in the next section.

\subsection{RQ3: Factors influencing eating behaviors}\label{sec:results:influences}
Ordinary least squares (OLS) regression was used to assess (1) the influences of sociodemographic factors on the journaling behaviors and (2) the influences of sociodemographic and journaling factors on the caloric and diet-based behavioral measures. To that end, we built 17 regression models in which each predicted variable corresponds to each behavioral measure. For predictor variables, we included gender (dummy coded 1 for male and 0 for female), age, social connections (logarithmic scale), US region (dummy coded into 3 variables using Midwest as a reference category), global region (dummy coded 1 for non-US and 0 for US), recording days, and lapsing frequency.

Table \ref{tbl:regression} displays the predictor variables and their coefficients ($\beta$) from 17 OLS regression models. The values of adjusted $R^2$ of all the models vary from 0.002 - 0.219, which are fairly low. As the predicted variables are derived from a long-term (up to 6 months) consumption data, we expect the predictors of the regression models to modestly explain a small portion of variance of the predicted variables as they do not take into account the temporal variability of the behaviors. This also points to the fact that there are many other factors which could potentially influence these behaviors. In what follows, we summarize the influences of different factors on all behaviors in the order of importance (by absolute coefficient values and total number of behavioral measures influenced).

\textbf{Gender:} Overall, gender appears to be the most important factor in influencing most of the behaviors in the study. Being male will substantially change the value of most behavioral measures compared to being female after controlling for other variables. Specifically, gender significantly influences 14 of 17 different behavioral measures. Amongst all predictor variables, gender has the highest predictive power on 7 behavioral measures, such as recording days, lapsing frequency, and caloric intake, and the relatively high predictive power (the top-3 highest coefficients) on 7 other measures, such as FV intake, FV intake frequency, and poultry intake frequency.


\textbf{Lapsing frequency:} Next, lapsing frequency is the second most influential factor of eating behaviors after gender. Specifically, it has a relatively high predictive power on 10 behavior measures, such as caloric intake, FV intake, and FV intake frequency. Interestingly, it has adverse relationships with most eating behaviors as indicated by negative coefficients $\beta$. In some cases, the negative relationships seem counter-intuitive, e.g., an increase in lapsing frequency decreases median caloric intake and intakes of unhealthy diets. This could be partially explained by the fact that some food journalers may be more incline to record less and less diary entries before they temporarily stop journaling \cite{Weber2016}. For some, this may positively indicate that they have already achieved their self-tracking goals -- an example of \textit{successful abandonment} \cite{Clawson2015}. In a recent study, De Choudhury et al. \cite{DeChoudhury2017} concluded that this phenomenon should not be common and long-term food journalers are more likely to complete their diary entries when they choose to record a journal.


\textbf{Regions of residence:} Regions of residence have a substantial influence on journaling and eating behaviors, particularly global regions (US vs. Non-US). After adjusting for other factors, being outside the US will considerably affect most behavioral measures compared to being in the US. 13 of 17 behavioral measures are significantly influenced by global regions. Furthermore, it has a relatively high predictive power on 12 behavioral measures, such as caloric intake, FV intake, and FV intake frequency. More importantly, it is one of the two significant predictors of fish intake. Next, US regions of residence (Northeast, South, and West) significantly influence 10 of 17 behavioral measures. Amongst the behavioral measures, it is a relatively high predictor of red and processed meat intake (Northeast), red and processed meat intake frequency (Northeast), sugary drink intake (Northeast), recording days (South), red and processed meat intake (South), poultry intake (South), sugar drink intake (South), soft drink intake frequency (South), and soft drink intake (West).


\textbf{Age:} Age is not as predictive of the healthy eating and journaling behaviors as gender, lapsing frequency, and regions of residence. Even though 13 of 17 behaviors are influenced by age, its effects ($\beta$) on these behaviors are fairly modest compared to many predictor variables. For example, a one year increase in age will increase the median red and processed meat intake by 0.004 servings after adjusting for other variables.

\textbf{Recording days:} Recording days is one of the least predictive factors of most dietary behaviors. A one day increase in recording days will marginally change most behavioral measures after controlling for other factors. Only half of the behavioral measures (8 of 16) are significantly influenced by recording days. In addition, it is the second worst predictor of 7 influenced measures, such as caloric intake, FV intake frequency, and red and processed meat intake. Interestingly, it is one of the two significant predictors of fish intake.


\textbf{Social connections:} Interestingly, social connection seems to be the least influential factor of most behavioral measures. That is, it has the least predictive power on 10 influenced behavior measures, such as recording days, lapsing frequency, and caloric intake. For instance, a 1\% increase in social connections will only increase the number of recording days by 0.007725 days after adjusting for other factors.


\textbf{Key insights:} Between the two journaling factors considered in this study, journaling persistence is more predictive of the healthy eating behaviors than journaling duration. Moreover, the journaling duration is one of the least influential factors of behaviors compared to other sociodemographic factors. More importantly, many sociodemographic factors, especially gender and regions of residence, still play a more critical role in determining the healthy eating behaviors of food journalers than the journaling factors. Contrary to its contribution in the weight loss outcomes, the results show that the journaling duration is a marginal determinant of the healthy eating behaviors.

\section{Discussion}

Millions of people use mobile food journals as a tool for tracking their caloric intake in order to achieve specific health goals, such as losing weight and living a healthy lifestyle. Prior findings have shown that food journals who actively record what they ate tend to meet their caloric goals \cite{Weber2016, DeChoudhury2017} and lose more weights \cite{Burke2012}. However, beyond managing caloric intake and weight, we found several lapses in the food journalers' healthy eating behaviors, e.g., low consumption of fruits and vegetables and fish, and high consumption of sugary drinks and soft drinks, preventing them from fully achieving their healthy lifestyle goal. Compared to the sociodemographic factors, the journaling duration has the least amount of influence on most eating behaviors. Most journalers' healthier food choices are significantly influenced by their gender roles and environmental contexts, suggesting that their food choices are still largely subconscious \cite{Wansink2006}. Overall, our study helps to further investigate the claim about the journaling effect on healthier food choices \cite{Cordeiro2015a, Cordeiro2015, Epstein2016} using a large-scale data. From the public health perspective, we find that their effectiveness in helping people achieve healthy eating behaviors as defined by evidence-based outcomes are marginal at best. Based on our findings, it appears that the calorie counting aspect of food journals is not sufficient and potentially unhelpful in facilitating sustained health behavior change. By exclusively focusing on caloric and weight control, food journals may unintentionally mislead individuals into pursuing caloric management as the only health goal and disregarding the physiological and metabolic effects of different types of diets on health and well-being. Our study also confirms findings from past study \cite{Cordeiro2015a}, suggesting that there exists a mismatch between food journalers' preconceived notions of health eating and the energy-centric design of mobile food journals, inadvertently leading to negative nudges.

Our findings have several implications to the designs of mobile food journals or other forms of \textit{mHealth} (mobile health) interventions that can better facilitate healthy eating behaviors. First, from the goal setting perspective \cite{Munson2012}, a variety of \textit{behavioral goals} could be suggested or posed as daily food challenges \cite{Epstein2016} to individuals by learning from their past food journal data. The aim is to: (1) supplement the existing caloric and nutrient intake goals; (2) provide a well-defined and quantifiable steps to help people achieve the healthy lifestyle goals, such as high daily FV intake; and (3) improve behavioral compliance and self-efficacy. Next, the goal-setting mechanism could incorporate individuals' backgrounds and experiences such that the behavioral goals could be dynamically and incrementally adjusted to suit them. Next, the goal-setting mechanism could be designed to focus more on targeting whole food consumption (e.g., \textit{consuming at least 5 servings of \textbf{fruits} and \textbf{vegetables} a day}) than isolated nutrients and constituents (e.g., \textit{consuming at least 30 grams of \textbf{fiber} per day}). This could effectively help educate individuals about the importance of food synergy and promote the idea of dietary variety \cite{Jacobs2009}. Lastly, behavioral interventions tailored to individuals' sociodemographic backgrounds could be introduced to mobile food journals. The aim is to identify individuals who are highly susceptible to certain eating behavior lapses and provide them with additional guidance and actions relevant to their current goals. For instance, male users, who are relatively prone to infrequent fruit and vegetable consumption, could receive more targeted notification messages designed to remind and persuade them to meet their FV goals. Additionally, other complementary sources of data, e.g., social media \cite{DeChoudhury2017}, could be included to improve the adaptability and effectiveness of the interventions.

\subsection{Limitations}
Due to the self-reported and user-contributed natures of data, it is difficult to verify the accuracy of the MFP data \cite{DeChoudhury2017, Weber2016, Cordeiro2015}. These issues could bias the estimations of portion sizes and nutrient intakes for various diet types, especially high-sugar diets which can be deliberately omitted due to guilt \cite{Cordeiro2015}. Next, external data about the dietary consumption of the general populace are generally collected through traditional dietary assessment instruments, such as a Food Frequency Questionnaire (FFQ), which are more susceptible to recall bias than daily food journals. Higher prices of healthier food choices, such as fish, may also affect the healthy eating patterns. Moreover, since it is not possible to control for food journalers' personal belief and perception about food and nutrition from the MFP data, the effects (or lack thereof) of the journaling practice may be confounded by the differences in healthy eating perception. Although most journalers generally hold the views of healthy eating which are consistent with evidence-based recommendations \cite{Cordeiro2015a}, some may choose to follow a specific dietary regimen, e.g., vegetarian diet, ketogenic diet, etc., for various reasons. Nevertheless, by allowing some noise in the data, our findings can be fairly compared to epidemiological studies'. Lastly, since our findings are obtained from online observational data, therefore they are limited in determining the causal associations between sociodemographic and journaling factors and healthy eating behaviors.




\section{Conclusion}
In this study, we investigated the healthy eating behaviors of MyFitnessPal food journalers. Despite the claim about the benefit of journaling in promoting healthier choices, we found that most food journalers did not eat more healthful diets than the general public. First, much of their dietary consumption did not meet the daily recommended intakes of healthy and unhealthy food sources. Next, their dietary patterns were not as uniform as we initially expected and the distinct patterns mostly resembled those of the general populace who may be less health conscious. Moreover, journaling duration, which was previously shown to be associated with improved weight loss outcomes, appeared to have a marginal influence on the healthy eating behaviors, whereas gender, lapsing frequency, and regions of residence are much more predictive of the healthy eating outcomes.


\section{Acknowledgement}
This research is supported by the National Research Foundation, Prime Minister's Office, Singapore under its International Research Centres in Singapore Funding Initiative.

\balance{}

\bibliographystyle{ACM-Reference-Format}
\bibliography{main}

\end{document}